\documentstyle[prl,aps,floats,epsf,color]{revtex}
\begin{document}
\baselineskip=12pt
\def\be{\begin{equation}}
\def\ee{\end{equation}}
\def\bea{\begin{eqnarray}}
\def\eea{\end{eqnarray}}
\def\E{{\rm e}}
\def\bearst{\begin{eqnarray*}}
\def\eearst{\end{eqnarray*}}
\def\peleven{\parbox{11cm}}
\def\peffec{\peight{\bearst\eearst}\hfill\peleven}
\def\pspace{\peight{\bearst\eearst}\hfill}
\def\ptwelve{\parbox{12cm}}
\def\peight{\parbox{8mm}}
\newcommand{\fixme}[1]{\footnote{\color{red}Fixme: #1}}
\twocolumn[\hsize\textwidth\columnwidth\hsize\csname@twocolumnfalse\endcsname

\title
{Fractal Analysis of River Flow Fluctuations }

\author{M. Sadegh Movahed$^{1,2}$, Evalds Hermanis$^{3}$}
\address{$^{1}$Department of Physics, Shahid Beheshti University, Evin, Tehran 19839, Iran}
\address{$^{2}$ Institute for Studies in theoretical Physics and Mathematics, P.O.Box 19395-5531,Tehran, Iran}
\address{$^{3}$ Laboratory Vide,  "Br\={\i}vi\c{n}i", Birzgale district, Ogre region, Latvia, LV-5033}

\vskip 1cm

 \maketitle


\begin{abstract}
 We use some fractal analysis methods to study river flow fluctuations.
 The result of the Multifractal Detrended Fluctuation Analysis (MF-DFA) shows that there are two crossover
timescales at $s_{1\times}\sim12$ and $s_{2\times}\sim130$ months in
the fluctuation function. We discuss how the existence of the
crossover timescales are related to a sinusoidal trend. The first
crossover is due to the seasonal trend and the value of second ones
is approximately equal to the well known cycle of sun activity.
Using Fourier detrended fluctuation analysis, the sinusoidal trend
is eliminated. The value of Hurst exponent of the runoff water of
rivers without the sinusoidal trend shows a long range correlation
behavior. For the Daugava river the value of Hurst exponent is
$0.52\pm0.01$ and also we find that these fluctuations have
multifractal nature. Comparing the MF-DFA results for the remaining
data set of Daugava river to those for shuffled and surrogate
series, we conclude that its multifractal nature is almost entirely
due to the broadness of probability density function.
\end{abstract}
\hspace{.3in}
\newpage
]
\section{Introduction}
Interpretation and estimation of climate change has been one of the
main research areas in science \cite{1,2,3}. The climate system
often exhibits irregular and complex behavior. Although the climate
system is driven by the well-defined seasonal periodicity, it is
also a subject to unpredictable perturbations which can lead to
extreme climate events. Indeed the climate is a dynamical system
influenced by immense factors, such as solar radiation or the
topography of the surface of the solid earth, etc. All factors that
control the trajectory of climate have enormously large phase space,
thus we have to analysis it with stochastic tools. Several recent
statistical studies have shown that a remarkably wide variety of
natural systems display fluctuations that may be characterized by
long-range power-law correlations. Such correlations hint toward
fractal geometry of the underlying dynamical system. Existence and
determination of power-law correlations would help to quantify the
underlying process dynamics \cite{val03,livina03}.

The analysis of river flows has a long history, nevertheless some
important issues have been lost. Here, we study one component of the
climate system, the river flux, by using the novel approach in the
fractal analysis like Detrended Fluctuation Analysis,
Fourier-Detrended Fluctuation Analysis and  Scaled Windowed Variance
Analysis Methods. The statistical and fractal analysis of river
flows should be an important issue in the geophysics and
hydrological systems to recognize the influence of environmental
conditions and to detect relative effects. A set of most important
results which can be given by using statistical tools is as follows:
a concept of scale self-similarity for the topography of Earth's
surface \cite{Mand1}, the hydraulic-geometric similarity of river
system and floods forced by the heavy rain \cite{Schmitt,bor}, etc.
 Already more than half a
century ago the engineer Hurst found that runoff records from
various rivers exhibit 'long-range statistical
dependencies'\cite{Hurst51}. Later, such long-term correlated
fluctuation behavior has also been reported for many other
geophysical records including precipitation data
\cite{Mand1,hurst65,Feder}. These original approaches exclusively
focused on the absolute values or the variances of the full
distribution of the fluctuations, which can be regarded as the first
and second moments of detrended fluctuation analysis
\cite{Mand1,Hurst51,hurst65,Matsoukas}. In the last decade it has
been realized that a multifractal description is required for a full
characterization of the runoff records \cite{Tessier,Pandey}. This
multifractal description of the records can be regarded as a
'fingerprint' for each station or river, which, among other things,
can serve as an efficient non-trivial test bed for the
state-of-the-art precipitation-runoff models.

River flow can be characterized by several general features. As a
result of the periodicity in precipitation, river flow has also
strong seasonal periodicity. The seasonal cycle of river flow is
asymmetric; i.e., river flow increases rapidly (usually during late
winter and spring) and decreases gradually (toward the end of the
autumn). The fluctuations in river flow are large for large river
flow and small for small river flow \cite{livina03}. It is important
to note that unlike other climate components, river flow may has a
direct impact of human activity, like damming, use of river water
for agriculture, etc., a fact which makes the river flow data more
difficult to study. The fluctuations in river flow are of special
interest since they are directly linked to floods and droughts.
There are several interesting characteristics of river flow
fluctuations: (i) the river flow fluctuations have power law tails
in the probability distribution \cite{mur,kroll}, (ii) the river
flow fluctuations are long-range correlated \cite{Hurst51,pell,bun},
and (iii) river flow fluctuations are multifractal
\cite{Tessier,Pandey,kante03}. The scaling laws may improve the
statistical prediction of extreme changes in river flow
\cite{bunde1}. Recently connection between volatility and
nonlinearity has originally been established by
\cite{Ashkenazy01,Ashkenazy03}, the degree of non-linearity has been
checked using the volatility series, also a simple model of river
fluctuations has been determined \cite{val03,livina03}. More
recently the annual runoff for the Ukrainian and Moldavia's rivers
and reveal scale invariance for distribution of this variable have
been investigated by using statistical parameters such as arithmetic
average, coefficients of variation, skewness, and auto-correlation
\cite{loboda}. In all of the previous researches, the contribution
sinusoidal trends on the creation of crossover in the results of
fractal analysis and the multifractal nature have been lost. The
effect of nonstationarity on the detrended fluctuation analysis has
been investigated \cite{kunhu}. In addition, the effects of periodic
trends on the fractal scaling properties of a time series have been
investigated moreover in some paper a relation between amplitude and
the period of the periodic trends and the existence of crossover in
the Detrended Fluctuations scaling function  have been demonstrated
\cite{kunhu}. So the main purpose of this paper are the
investigation of the effects of seasonal trend on the multifractal
analysis of flow fluctuations and determination of the source of
multifractality in data . For completeness of this investigation and
to get the deep insight of the contribution of sun activity in the
statistical properties of river flow, we compare the recent fractal
analysis results of sunspots \cite{sa06} with current analysis. The
Sunspot number was collected by the Sunspot Index Data Center (SIDC)
\cite{data}. Due to the stochastic nature of river flow, it is
probable that the sun activity may affects on the duct of river, so
to demonstrate the presence of any correlations we compare results
of river flow and sun activity extracted by various fractal
analysis. It was well-known that the statistical properties of every
rivers depend on very important reasons which affect on flow
fluctuations, so one can not expect that the sun activity has a same
reasonable effect on different rivers.

In addition we would like to characterize the complex behavior of
the monthly runoff for the Daugava river fluctuations through the
computation of the signal parameters - scaling exponents - which
quantifies the correlation exponents and multifractality of the
signal. We investigate the correlation behavior of duct river time
series which is governed by power-law.

\begin{figure}[t]
\epsfxsize=8truecm\epsfbox{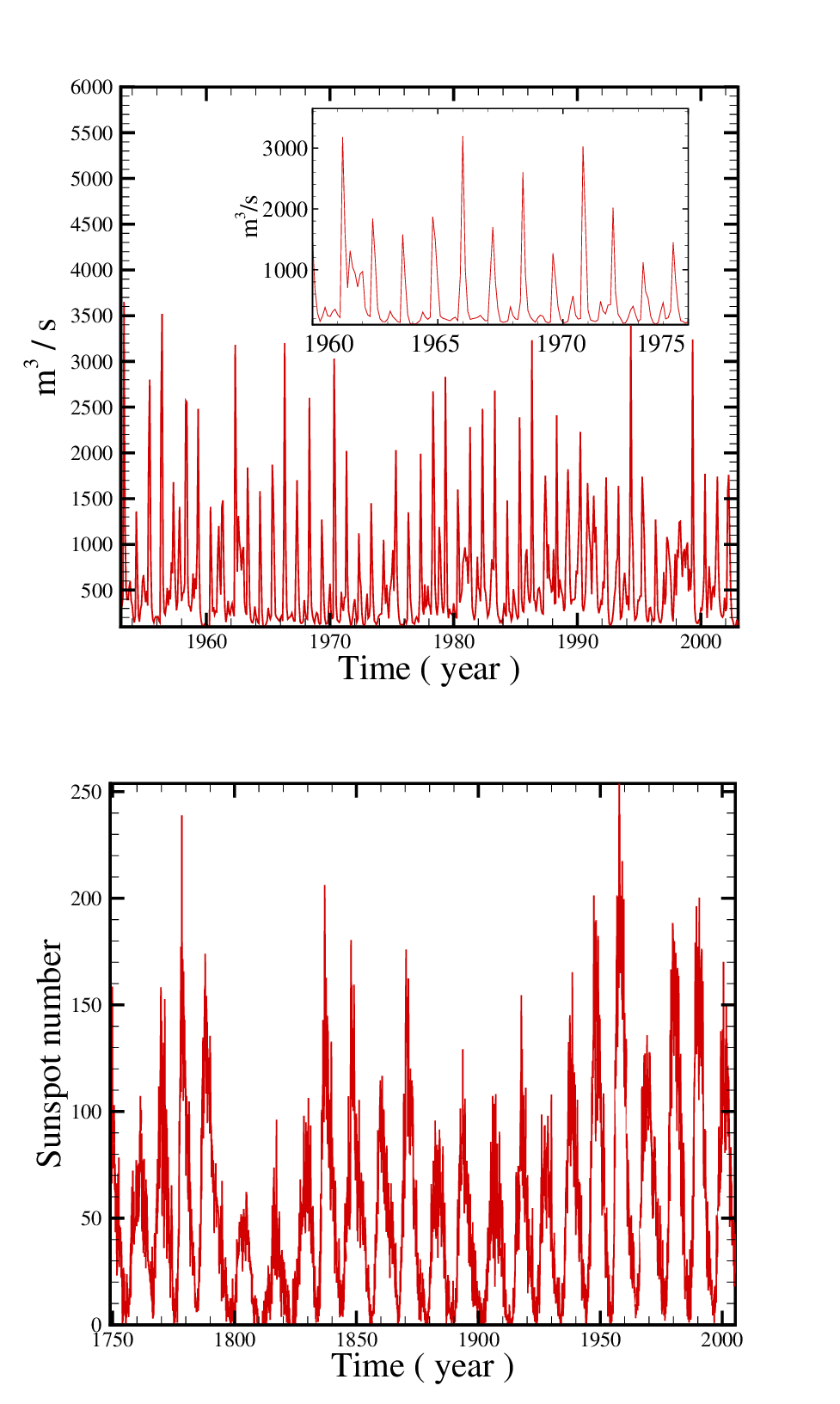} \narrowtext \caption{ Observed
flux series of Daugava river (upper panel) and Sunspot number (lower
panel) as a function of time. } \label{fig1}
 \end{figure}

The original Daugava river data source is Latvian Environmental
Geological and Meteorological Agency database. They describe water
flow through hydroelectric power station near \c{K}egums, Latvia.
Dimension of the data is m$^3$/s. Other data which are used here are
from National Water Information System: Web Interface
\cite{data_river}. As shown in the upper panel of Figure \ref{fig1},
the duct water of Daugava river series has a sinusoidal trend, with
a dominant frequency. These trends should involve the seasonal and
other physical reasons in natural phenomenon. Because of the
complexity nature of river flow series, and due to the finiteness of
the available data sample, we should apply some methods which are
insensitive to non-stationarities like trends.
To eliminate the effect of sinusoidal trend, we apply the Fourier
Detrended Fluctuation Analysis (F-DFA) \cite{na04,chi05}. After
elimination of the trend we use the Multifractal Detrended
Fluctuation Analysis (MF-DFA) to analysis the data set. The MF-DFA
 methods are the modified version of detrended fluctuation
analysis (DFA) to detect multifractal properties of time series. The
detrended fluctuation analysis (DFA) method introduced by Peng et
al. \cite{Peng94} has became a widely-used technique for the
determination of (multi-) fractal scaling properties and the
detection of long-range correlations in noisy, nonstationary time
series \cite{kunhu,Peng94,murad,physa,kunhu1}. It has successfully
been applied to diverse fields such as DNA sequences
\cite{Peng94,dns}, heart rate dynamics \cite{herz,Peng95,PRL00},
neuron spiking \cite{neuron}, human gait \cite{gait}, long-time
weather records \cite{wetter}, cloud structure \cite{cloud}, geology
\cite{malamudjstatlaninfer1999}, ethnology \cite{Alados2000},
economical time series \cite{economics,Ivanov04}, solid state
physics \cite{fest}, sunspot time series \cite{sa06} and cosmic
microwave background radiation \cite{sacmb}.

This paper is organized as follows:  In Section II we describe the
MF-DFA, F-DFA and Scale Windowed Variance (SWV) methods in detail
and show that the scaling exponents determined via the MF-DFA method
are identical to those obtained by the standard multifractal
formalism based on partition functions. We eliminate the sinusoidal
trend via the F-DFA technique in Section III and investigate the
multifractal nature of the remaining fluctuation, we use certain
fractal analysis approaches such as, the Multifractal Detrended
Fluctuation Analysis (MF-DFA), and Scaled Windowed Variance (SWV) to
analysis the data set, The DFA result of sun activity and river
flows are compared together. In Section IV, we examine the source of
multifractality in duct water of Daugava river data by comparison
the MF-DFA results for remaining data set to those obtained via the
MF-DFA for shuffled and surrogate series.
Section V closes with a discussion of the present results.\\

\section{Fractal Analysis Methods}

In this section we introduce three methods to investigate the
fractal properties of stochastic processes.
\subsection{Multifractal Detrended Fluctuation Analysis}
The simplest type of the multifractal analysis is based upon the
standard partition function multifractal formalism, which has been
developed for the multifractal characterization of normalized,
stationary measurements \cite{feder88,barabasi,peitgen,bacry01}.
Unfortunately, this standard formalism does not give correct results
for nonstationary time series that are affected by trends or that
cannot be normalized. Thus, in the early 1990s an improved
multifractal formalism has been developed, the wavelet transform
modulus maxima (WTMM) method \cite{wtmm}, which is based on the
wavelet analysis and involves tracing the maxima lines in the
continuous wavelet transform over all scales. The other method, the
multifractal detrended fluctuation analysis (MF-DFA), is based on
the identification of scaling of the $q$th-order moments depending
on the signal length and is generalization of the standard DFA using
only the second moment $q=2$.

The MF-DFA does not require the modulus maxima procedure in contrast
WTMM method, and hence does not require more effort in programming
and computing than the conventional DFA. On the other hand, often
experimental data are affected by non-stationarities like trends,
which have to be well distinguished from the intrinsic fluctuations
of the system in order to find the correct scaling behavior of the
fluctuations. In addition very often we do not know the reasons for
underlying trends in collected data and even worse we do not know
the scales of the underlying trends, also, usually the available
record data is small. For the reliable detection of correlations, it
is essential to distinguish trends from the fluctuations intrinsic
in the data. Hurst rescaled-range analysis \cite{hurst65} and other
non-detrending methods work well if the records are long and do not
involve trends. But if trends are present in the data, they might
give wrong results. Detrended fluctuation analysis (DFA) is a
well-established method for determining the scaling behavior of
noisy data in the presence of trends without knowing their origin
and shape \cite{Peng94,Peng95,fano,allan,buldy95}. Also DFA scaling
results are not immune to different trends and to different
artifacts such as spikes, missing segments of data etc.
\cite{kunhu1,chen05,xu05}.

In spite of many abilities of this method, in some cases is
encountered with problem and gives wrong results. DFA method can
only determine positive Hurst exponent, $H$, and gives an inaccurate
results for strongly anti-correlated record data when $H$ is close
to zero. To avoid this situation, in such case one should use the
integrated data. This signal is so-called double profiled data set.
The corresponding Hurst exponent using this way is $H=\bar{H}-1$,
here $\bar{H}$ is derived from DFA method for the double profiled
signals \cite{sa06,bun02}. According to the recent exploration in
ref. \cite{physa}, a deviation in the DFA results occurs in very
short records and in the small regime of data in each window
mentioned in the forthcoming subsection. The modified version of
MF-DFA will be used for these cases \cite{physa}. In this paper also
we will introduce this method and apply to infer correct exponent
for river flow fluctuations.


\subsection{Description of the MF-DFA}

The modified multifractal DFA (MF-DFA) procedure consists of five
steps\cite{bun02} \fixme{The Multifractal Detrended Fluctuation Analysis (MF-DFA) method has been introduced by J. W. Kantelhardt et. Al. in \cite{bun02}. In this research, we only applied this approach in our work. In this part of our paper we used the same text that the authors of \cite{bun02} used previously.}.  The first three steps are essentially identical to the
conventional DFA procedure (see e.~g.
\cite{kunhu,Peng94,murad,physa,kunhu1}). Suppose that $x_k$ is a
series of length $N$, and that this series is of compact support,
i.e. $x_k = 0$ for an insignificant fraction of the values only.

\noindent $\bullet$ {\it Step 1}: Determine the ``profile''

\begin{equation}
Y(i) \equiv \sum_{k=1}^i \left[ x_k - \langle x \rangle \right] ,
 \qquad i=1,\ldots,N.
\label{profile}
\end{equation}
Subtraction of the mean $\langle x \rangle$ is not compulsory, since
it would be eliminated by the later detrending in the third step.

\noindent $\bullet$ {\it Step 2}: Divide the profile $Y(i)$ into
$N_s \equiv {\rm int}(N/s)$ non-overlapping segments of equal
lengths $s$. Since the length $N$ of the series is often not a
multiple of the considered time scale $s$, a short part at the end
of the profile may remain.  In order not to disregard this part of
the series, the same procedure is repeated starting from the
opposite end.  Thereby, $2 N_s$ segments are obtained altogether.

\noindent $\bullet$ {\it Step 3}: Calculate the local trend for each
of the $2 N_s$ segments by a least-square fit of the series.  Then
determine the variance
\begin{equation} F^2(s,\nu) \equiv {1 \over s} \sum_{i=1}^{s}
\left\{ Y[(\nu-1) s + i] - y_{\nu}(i) \right\}^2, \label{fsdef}
\end{equation}
for each segment $\nu$, $\nu = 1, \ldots, N_s$ and
\begin{equation} F^2(s,\nu) \equiv {1 \over s} \sum_{i=1}^{s}
\left\{ Y[N - (\nu-N_s) s + i] - y_{\nu}(i) \right\}^2,
\label{fsdef2}
\end{equation}
for $\nu = N_s+1, \ldots, 2 N_s$.  Here, $y_{\nu}(i)$ is the fitting
polynomial in segment $\nu$.  Linear, quadratic, cubic, or higher
order polynomials can be used in the fitting procedure
(conventionally called DFA1, DFA2, DFA3, $\ldots$)
\cite{Peng94,PRL00}. Since the detrending of the time series is done
by the subtraction of the polynomial fits from the profile,
different order DFA differ in their capability of eliminating trends
in the series.  In (MF-)DFA$m$ [$m$th order (MF-)DFA] trends of
order $m$ in the profile (or, equivalently, of order $m - 1$ in the
original series) are eliminated.  Thus a comparison of the results
for different orders of DFA allows one to estimate the type of the
polynomial trend in the time series \cite{kunhu,physa}.

\noindent $\bullet$ {\it Step 4}: Average over all segments to
obtain the $q$-th order fluctuation function, defined as:
\begin{equation} F_q(s) \equiv \left\{ {1 \over 2 N_s}
\sum_{\nu=1}^{2 N_s} \left[ F^2(s,\nu) \right]^{q/2} \right\}^{1/q},
\label{fdef}\end{equation}
where, in general, the index variable $q$ can take any real value
except zero.  For $q=2$, the standard DFA procedure is retrieved.
Generally we are interested in how the generalized $q$ dependent
fluctuation functions $F_q(s)$ depend on the time scale $s$ for
different values of $q$.  Hence, we must repeat steps 2, 3 and 4 for
several time scales $s$.  It is apparent that $F_q(s)$ will increase
with increasing $s$.  Of course, $F_q(s)$ depends on the DFA order
$m$. By construction, $F_q(s)$ is only defined for $s \ge m+2$.

\noindent $\bullet$ {\it Step 5}: Determine the scaling behavior of
the fluctuation functions by analyzing log-log plots of $F_q(s)$
versus $s$ for each value of $q$. If the series $x_i$ are long-range
power-law correlated, $F_q(s)$ increases, for large values of $s$,
as a power-law,
\begin{equation} F_q(s) \sim s^{h(q)} \label{Hq}. \end{equation}
In general, the exponent $h(q)$ may depend on $q$.   For stationary
time series such as  fGn (fractional Gaussian noise), $Y(i)$ in Eq.
\ref{profile}, will be a fBm (fractional Brownian motion) signal,
so, $0<h(q=2)<1.0$ (see the appendix for more details). The exponent
$h(2)$ is identical to the well-known Hurst exponent $H$
\cite{Peng94,murad,feder88}. Also for a nonstationary  signal, such
as fBm noise, $Y(i)$ in Eq. \ref{profile}, will be a sum of fBm
signal, so the corresponding scaling exponent of $F_q(s)$ is
identified by $h(q=2)>1.0$ \cite{Peng94,eke02}. In this case the
relation between the exponents $h(2)$ and $H$ will be $H=h(q=2)-1$
(see appendix of \cite{sa06}). The exponent $h(q)$ is known as
generalized Hurst exponent. The auto-correlation function can be
characterized by a power law $C(s)\equiv\langle n_kn_{k+s} \rangle
\sim s^{-\gamma}$ with exponent $\gamma=2-2H$. Its power spectra can
be characterized by $S(\omega)\sim\omega^{-\beta}$ with frequency
$\omega$ and $\beta=2H-1$. In the nonstationary case, correlation
function is (see appendix for more details):
\begin{equation}
C(i,j)=\langle n_in_{j}\rangle\sim i^{2H}+j^{2H}-|i-j|^{2H}
\end{equation}
 and corresponding power spectrum scaling is $\beta=2H+1$
 \cite{sa06,Peng94,eke02}.

For monofractal time series, $h(q)$ is independent of $q$, since the
scaling behavior of the variances $F^2(s,\nu)$ is identical for all
segments $\nu$, and the averaging procedure in Eq.~(\ref{fdef}) will
just give this identical scaling behavior for all values of $q$. If
we consider positive values of $q$, the segments $\nu$ with large
variance $F^2(s,\nu)$ (i.~e. large deviations from the corresponding
fit) will dominate the average $F_q(s)$.  Thus, for positive values
of $q$, $h(q)$ describes the scaling behavior of the segments with
large fluctuations. For negative values of $q$, the segments $\nu$
with small variance $F^2(s,\nu)$ will dominate the average $F_q(s)$.
Hence, for negative values of $q$, $h(q)$ describes the scaling
behavior of the segments with small fluctuations \cite{note3}.

\subsubsection{ Relation to standard multifractal analysis}
For a stationary, normalized series the multifractal scaling
exponents $h(q)$ defined in Eq.~(\ref{Hq}) are directly related to
the scaling exponents $\tau(q)$ defined by the standard partition
function-based multifractal formalism as shown below \cite{bun02}\fixme{The Multifractal Detrended Fluctuation Analysis (MF-DFA) method has been introduced by J. W. Kantelhardt et. Al. in \cite{bun02}. In this research, we only applied this approach in our work. In this part of our paper we used the same text that the authors of \cite{bun02} used previously.}. Suppose that
the series $x_k$ of length $N$ is a stationary, normalized sequence.
Then the detrending procedure in step 3 of the MF-DFA method is not
required, since no trend has to be eliminated. Thus, the DFA can be
replaced by the standard Fluctuation Analysis (FA), which is
identical to the DFA except for a simplified definition of the
variance for each segment $\nu$, $\nu = 1, \ldots, N_s$. Step 3 now
becomes [see Eq.~(\ref{fsdef})]:
\begin{equation} F_{\rm FA}^2(s,\nu) \equiv [Y(\nu s) - Y((\nu-1) s)]^2.
\label{FAfsdef} \end{equation}
Inserting this simplified definition into Eq.~(\ref{fdef}) and using
Eq.~(\ref{Hq}), we obtain
\begin{equation} \left\{ {1 \over 2 N_s} \sum_{\nu=1}^{2 N_s}
\vert Y(\nu s) - Y((\nu-1) s) \vert^q \right\}^{1/q} \sim s^{h(q)}.
\label{FAfHq} \end{equation}
For simplicity we can assume that the length $N$ of the series is an
integer multiple of the scale $s$, obtaining $N_s = N/s$ and
therefore
\begin{equation} \sum_{\nu=1}^{N/s} \vert Y(\nu s) - Y((\nu-1) s)
\vert^q \sim s^{q h(q) - 1}. \label{MFA} \end{equation}
This corresponds to the multifractal formalism used e.~g. in
\cite{barabasi,bacry01}. In fact, a hierarchy of exponents $H_q$
similar to our $h(q)$ has been introduced based on Eq.~(\ref{MFA})
in \cite{barabasi}. In order to relate also to the standard textbook
box counting formalism \cite{feder88,peitgen}, we employ the
definition of the profile in Eq.~(\ref{profile}). It is evident that
the term $Y(\nu s) - Y((\nu-1) s)$ in Eq.~(\ref{MFA}) is identical
to the sum of the numbers $x_k$ within each segment $\nu$ of size
$s$. This sum is known as the box probability $p_s(\nu)$ in the
standard multifractal formalism for normalized series $x_k$,
\begin{equation} p_s(\nu) \equiv \sum_{k=(\nu-1) s +1}^{\nu s} x_k =
Y(\nu s) - Y((\nu-1) s).  \label{boxprob} \end{equation}
The scaling exponent $\tau(q)$ is usually defined via the partition
function $Z_q(s)$,
\begin{equation} Z_q(s) \equiv \sum_{\nu=1}^{N/s} \vert p_s(\nu)
\vert^q \sim s^{\tau(q)}, \label{Zq} \end{equation}
where $q$ is a real parameter as in the MF-DFA method, discussed
above. Using Eq.~(\ref{boxprob}) we see that Eq.~(\ref{Zq}) is
identical to Eq.~(\ref{MFA}), and obtain analytically the relation
between the two sets of multifractal scaling exponents,
\begin{equation} \tau(q) = q h(q) - 1. \label{tauH} \end{equation}
Thus, we observe that $h(q)$ defined in Eq.~(\ref{Hq}) for the
MF-DFA is directly related to the classical multifractal scaling
exponents $\tau(q)$.  Note that $h(q)$ is different from the
generalized multifractal dimensions
\begin{equation} D(q) \equiv {\tau(q) \over q-1} =
{q h(q)-1 \over q-1}, \label{Dq} \end{equation} that are used
instead of $\tau(q)$ in some papers.  While $h(q)$ is independent of
$q$ for a monofractal time series, $D(q)$ depends on $q$ in this
case. Another way to characterize a multifractal series is the
singularity spectrum $f(\alpha)$, that is related to $\tau(q)$ via a
Legendre transform \cite{feder88,peitgen},
\begin{equation} \alpha = \tau'(q) \quad {\rm and} \quad
f(\alpha) = q \alpha - \tau(q). \label{Legendre} \end{equation}
Here, $\alpha$ is the singularity strength or H\"older exponent,
while $f(\alpha)$ denotes the dimension of the subset of the series
that is characterized by $\alpha$. Using Eq.~(\ref{tauH}), we can
directly relate $\alpha$ and  $f(\alpha)$ to $h(q)$,
\begin{equation} \alpha = h(q) + q h'(q) \quad {\rm and} \quad
f(\alpha) = q [\alpha - h(q)] + 1.\label{Legendre2} \end{equation}

A H\"older exponent denotes monofractality, while in the
multifractal case, the different parts of the structure are
characterized by different values of $\alpha$, leading to the
existence of the spectrum $f(\alpha)$.

\subsection{Fourier-Detrended Fluctuation Analysis}

Many signals in the nature do not distinguished as monofractal
scaling behavior. In some cases, there exist one or more crossover
(time) scales $s_\times$ segregating regimes with different scaling
exponents e.g long range correlation for $s\ll s_{\times}$ and an
other type of correlation or uncorrelated behavior for $s\gg
s_{\times}$ \cite{kunhu,physa}. In other cases investigation of the
scaling behavior is more complicated. In the presence of different
behavior of various moments in the MF-DFA method, different scaling
exponents are required for different parts of the series
\cite{kunhu1}. Therefore one needs a multitude of scaling exponents
(multifractality) for a full description of the scaling behavior. A
crossover usually can arise from a change in the correlation
properties of the signal at different time or space scales, or can
often arise from trends in the data. To remove the crossover due to
a trend such as sinusoidal trends, Fourier-Detrended Fluctuation
Analysis (F-DFA) is applied. The F-DFA is a modified approach for
the analysis of low frequency trends added to a noise in time series
\cite{na04,chi05,koscielny98,koscielny98b}.

In order to investigate how we can remove trends having a low
frequency periodic behavior, we transform data record to Fourier
space, then we truncate the first few coefficient of the Fourier
expansion and inverse Fourier transform the series. After removing
the sinusoidal trends we can obtain the fluctuation exponent by
using the direct calculation of the MF-DFA. If truncation numbers
are sufficient, The crossover due to a sinusoidal trend in the
log-log plot of $F_q(s)$ versus $s$ disappears.

\subsection{Scaled Windowed Variance Analysis}
The Scaled Windowed Variance analysis was developed by Cannon et al.
(1997)\cite{eke02}. The profile of data, $Y(i)$, is divided into
$N_s \equiv {\rm int}(N/s)$ non-overlapping segments of equal
lengths $s$. Then the standard deviation is calculated within each
interval using the following relation
\begin{equation}
{\rm SWV}(s)=\left(\frac{1}{s}\sum_{i=1}^{s}[Y(i)-\langle
Y(s)\rangle]^2\right)^{1/2}.
\end{equation}
  The average standard deviation of all windows of length $s$ is computed. This
computation is repeated over all possible interval lengths. The
scaled windowed variance is related to $s$ by a power law
\begin{equation}\label{swv}
{\rm SWV}\sim s^H.
\end{equation}

\section{Analysis of data}

\begin{figure}[t]
\epsfxsize=8truecm\epsfbox{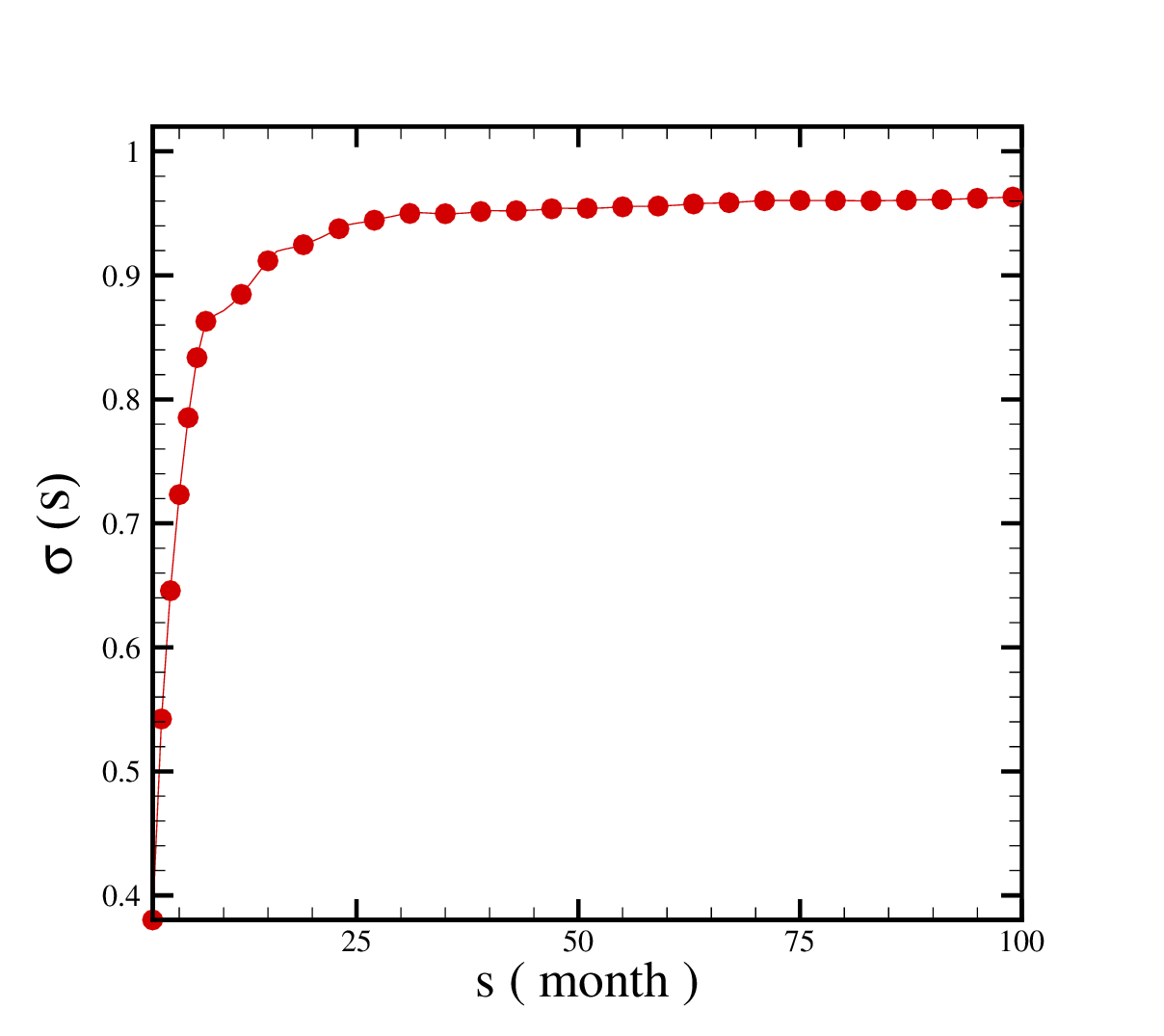} \narrowtext \caption{Behavior
of standard deviation of duct of water as a function of time scale.
It shows that this time series is stationary.} \label{fig2}
 \end{figure}

\begin{figure}[t]
\epsfxsize=8truecm\epsfbox{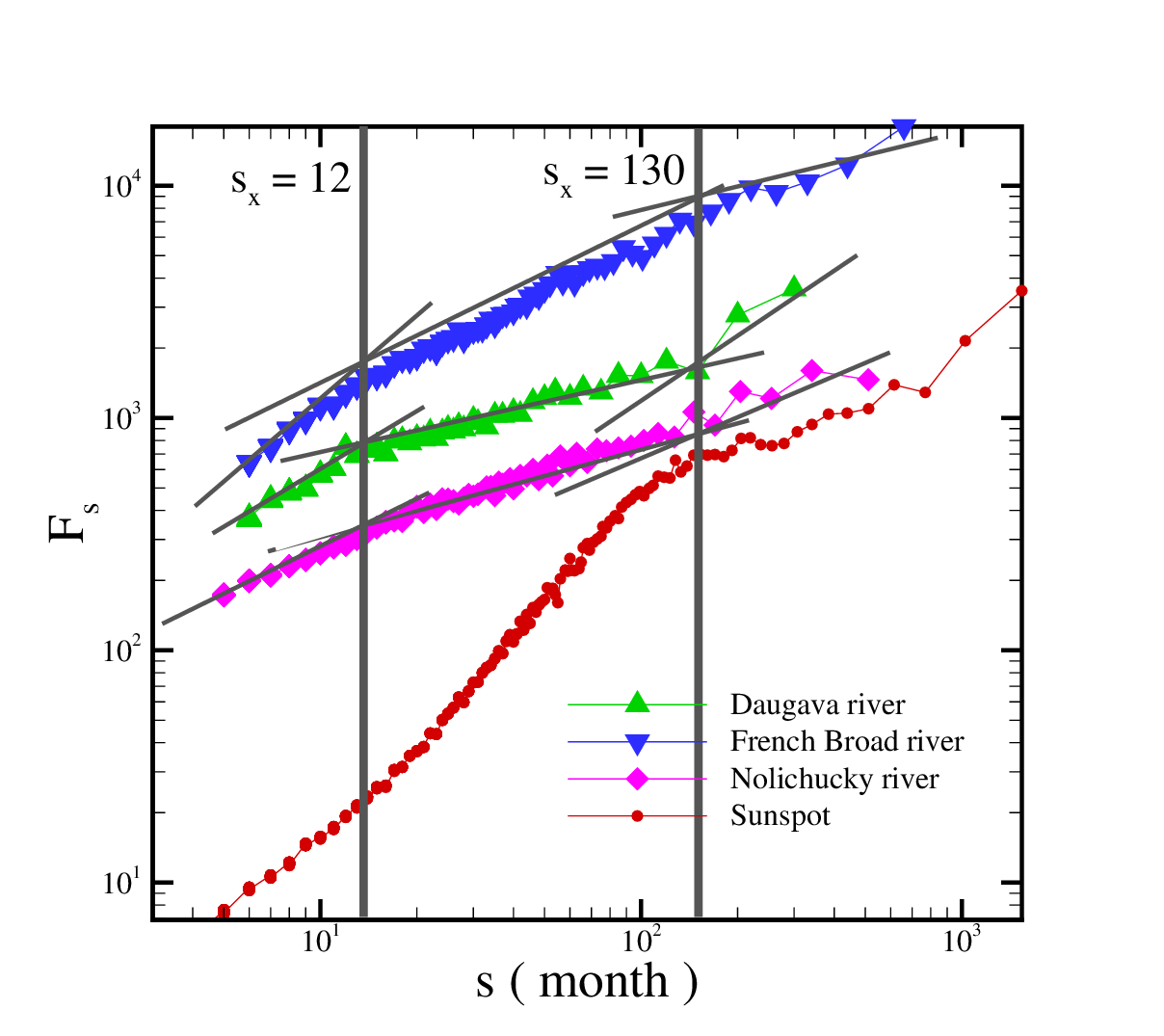} \narrowtext \caption{Crossover
behavior of log-log plot $F(s)$ versus $s$ for some river flows and
Sunspot series for $q=2.0$. For rivers  there are at least two
crossover time scales in plot of $F(s)$, at scales $s_{1\times}$ and
$s_{2\times}$.}
\label{fig3}
 \end{figure}

\begin{figure}[t]
\epsfxsize=8truecm\epsfbox{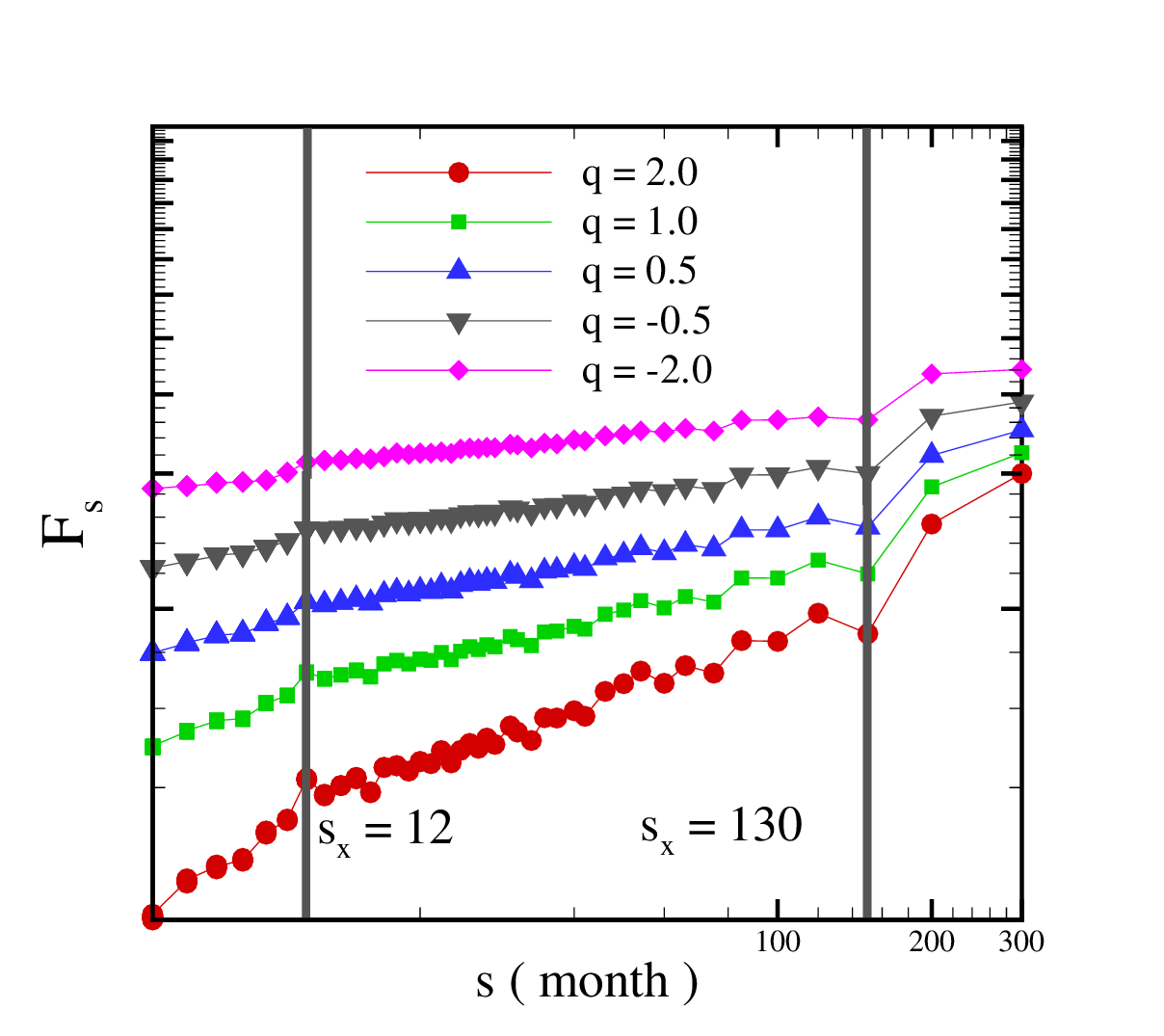} \narrowtext \caption{Crossover
behavior of log-log plot $F(s)$ versus $s$ for Daugava river flow
for various moments indicated on the figure. For river there are at
least two crossover time scales in plot of $F(s)$, at scales
$s_{1\times}$ and $s_{2\times}$. Vertical axis has arbitrary unite.}
\label{q}
 \end{figure}

As mentioned in section II, a spurious of correlations may be
detected if time series is nonstationarity, so direct calculation of
correlation behavior, spectral density exponent, fractal dimensions
etc., don't give the reliable results. It can be checked that the
runoff for Daugava river is or not nonstationary. One can verified
the non-stationarity property experimentally by measuring the
stability of the average and variance in a moving window for example
with scale $s$. Figure~\ref{fig2} shows the standard deviation of
Daugava flow signal versus scale $s$, is saturated. Let us determine
that whether the data set has a sinusoidal trend or not. According
to the MF-DFA1 method, Generalized Hurst exponents $h(q)$ in
Eq.~(\ref{Hq}) can be found by analyzing log-log plots of $F_q(s)$
versus $s$ for each $q$. Our investigation shows that there are at
least two crossover time scales $s_{\times}$ in the log-log plots of
$F_q(s)$ versus $s$ for every $q$'s. These two crossovers divide
$F_q(s)$ into three regions, as shown in Figure~\ref{fig3} ( for
instance we took $q=2$). Figure \ref{q} shows these crossover also
exist in the fluctuation function for various moments.  The
existence of these regions is due to the competition between noise
and sinusoidal trend \cite{kunhu}. For $s<s_{1\times}$, the noise
has the dominating effect . For $s_{1\times}<s<s_{2\times}$ the
sinusoidal (such as seasonal) trend dominates \cite{kunhu}. The
values of $s_{1\times}$ and $s_{2\times}$ are approximately equal to
$12$ and $130$ months, respectively. The first crossover is clearly
related to seasonal trend and the second ones is approximately equal
to the well known cycle of sun activity. This shows that in addition
of seasonal effect on the river flow, the sun activity strictly
affects on the feature and fractal properties of river flow
fluctuations. As shown in Figure \ref{fig3}, by comparing curves
from some rivers and Sunspot we can see a certain symmetry in the
form of curves. Points of inflection are placed closely, but angle
of bendings are opposite
(e.g. for Daugava river). 
This symmetry indicates a connection between the activity of sun and
flow of water in the rivers. In both cases sinusoidal tendency have
been found in the average part of the curves. Apparently the sun
activity governs the rivers.

\begin{figure}
\epsfxsize=8truecm\epsfbox{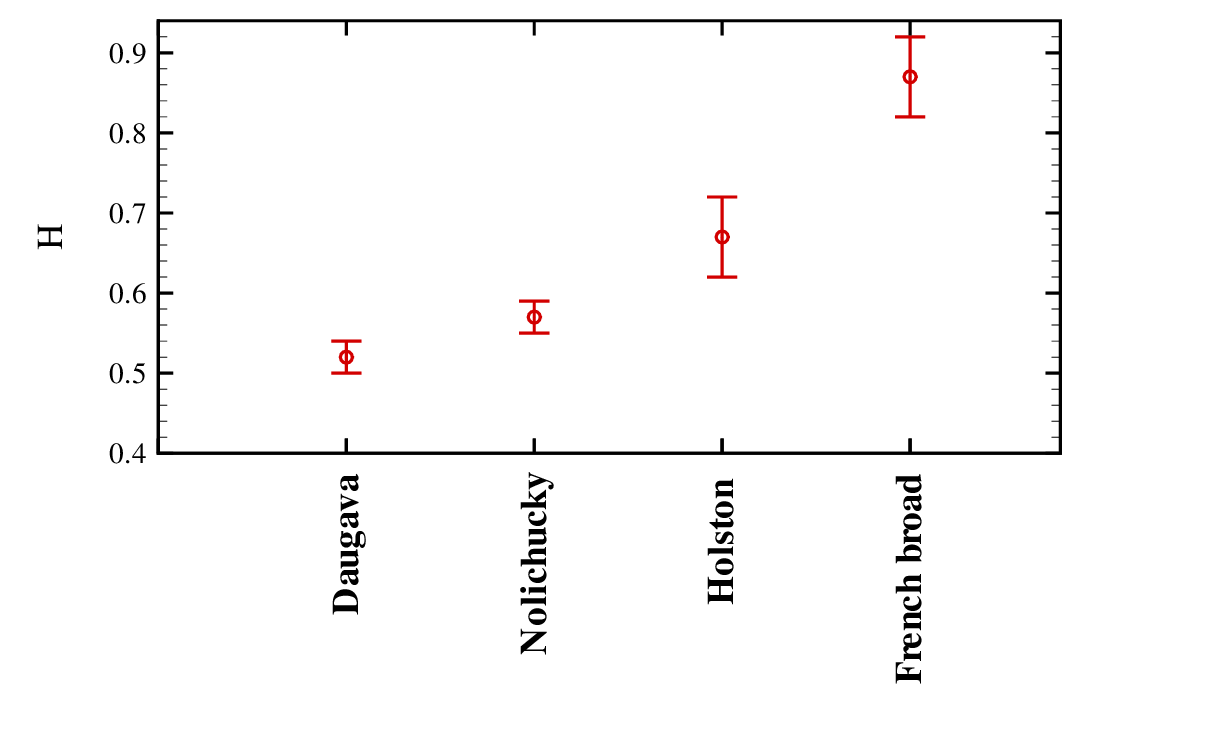} \narrowtext \caption{The values
of Hurst exponent for some famous rivers.} \label{hurstr}
 \end{figure}

 As mentioned before, for very small scales
$s<s_{1\times}$ the effect of the sinusoidal trend is not
pronounced, indicating that in this scale region the signal can be
considered as noise fluctuating around a constant which is filtered
out by the MF-DFA1 procedure. In this region the generalized DFA1
exponent for used rivers in this paper are listed in Figure
\ref{hurstr} 
, where confirms that the
process is a stationary process with long-range correlation
behavior.

\begin{figure}
\epsfxsize=8truecm\epsfbox{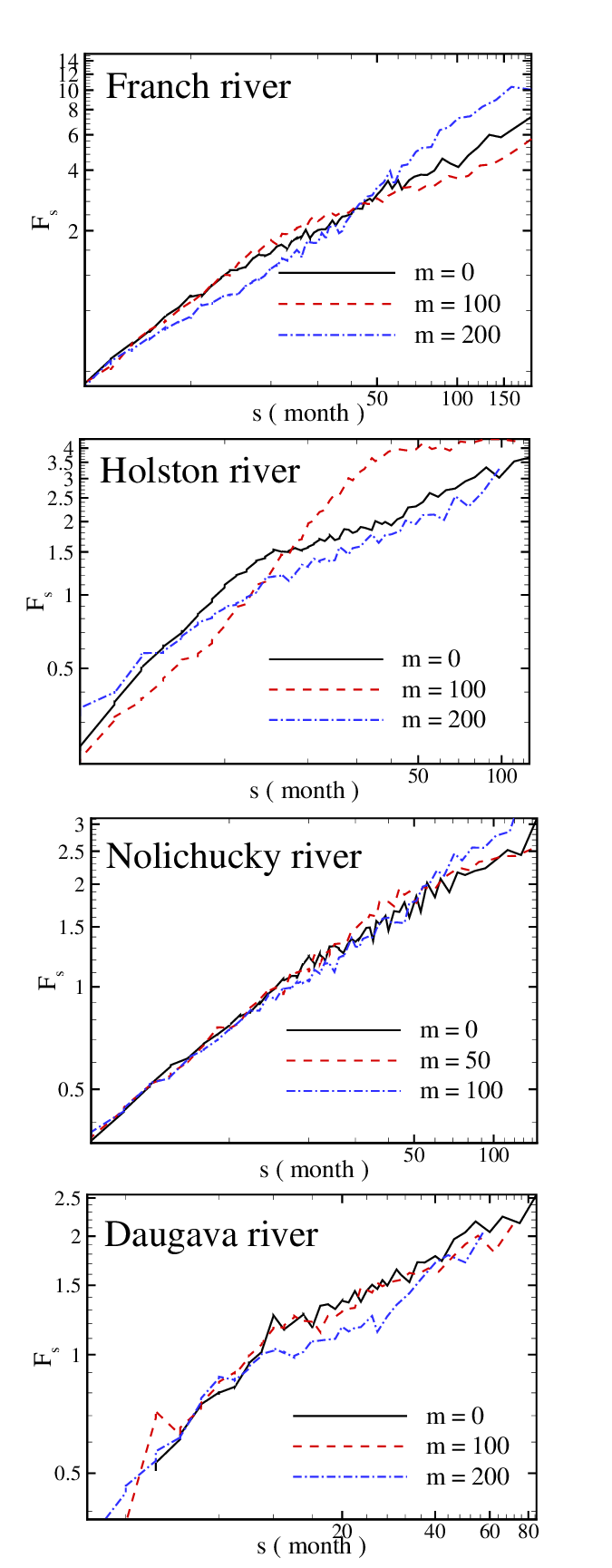} \vspace{5mm}\narrowtext
\caption{The MF-DFA1 functions $F_2(s)$ versus the time scale $s$ in
log-log plot. Original time series $m=0$, truncation of $m=50$,
$m=100$ and $m=200$ first terms.} \label{fig4}
 \end{figure}

 To cancel the sinusoidal trend in
MF-DFA1, we apply F-DFA method on the present data. We truncate some
of the first coefficients of the Fourier expansion of the river flow
fluctuations. According to Figure \ref{fig4}, for eliminating the
crossover scales, we need to remove approximately the first $200$
terms of the Fourier expansion. Then, by inverse Fourier
Transformation, the noise without sinusoidal trend is extracted (see
Figure \ref{fig4}) \cite{na04,chi05,koscielny98,koscielny98b}.

\begin{figure}[t]
\epsfxsize=8truecm\epsfbox{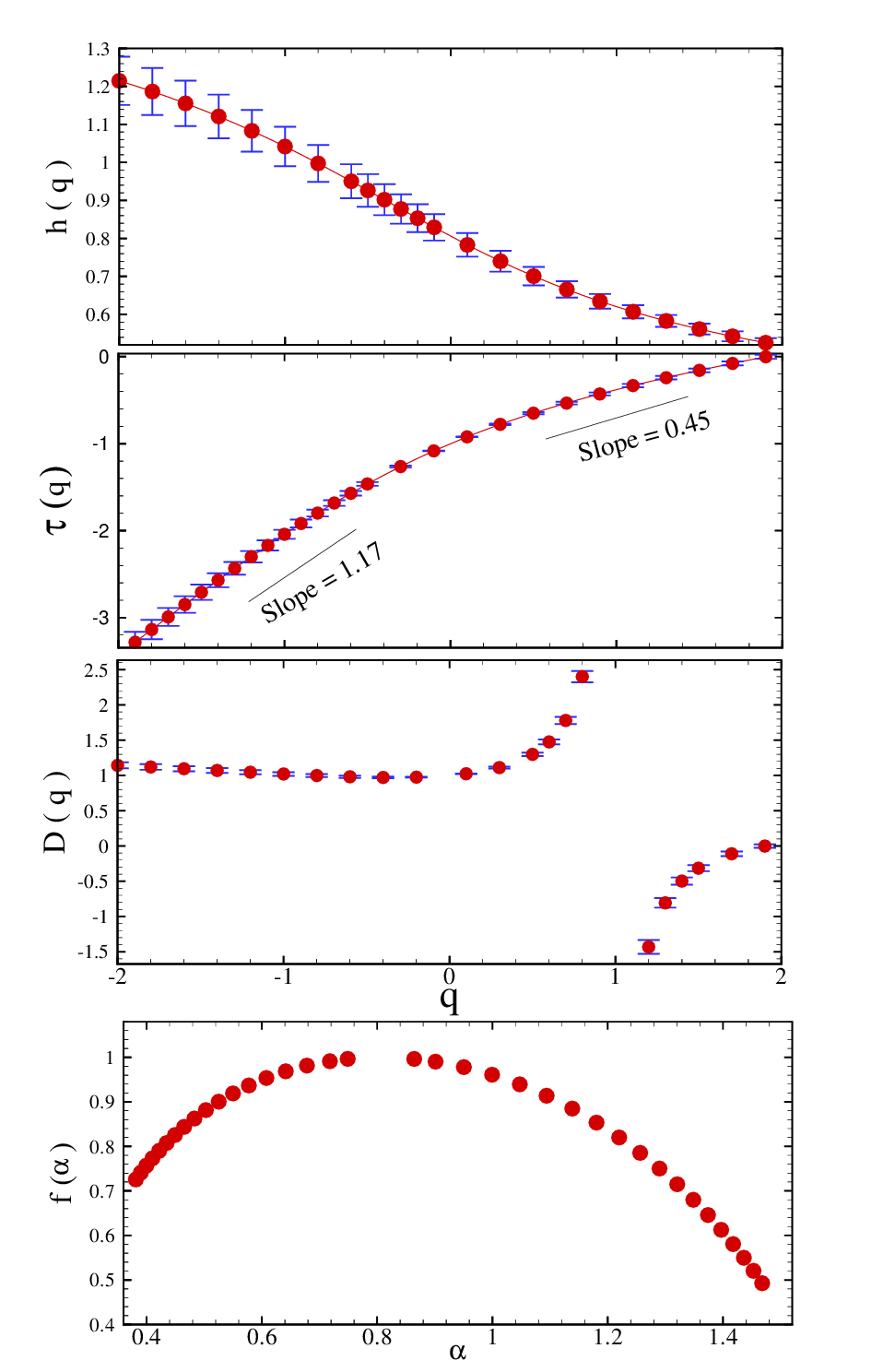} \narrowtext \caption{The $q$
dependence of the generalized Hurst exponent $h(q)$, the
corresponding $\tau(q)$, generalized multifractal dimension $D(q)$
and singularity spectrum $f(\alpha)$  are shown in the upper to
lower panel respectively for duct water of Daugava river series
without sinusoidal trend.} \label{fig5}
 \end{figure}

The MF-DFA1 results of the remanning new signal just for Daugava
river are shown in Figure \ref{fig5}. All exponent in this figure
are driven at as maximum scale as possible in the log-log plots of
$F_q(s)$ as function of $s$ (see Appendix for more details). The
duct water of Daugava river series is a multifractal process as
indicated by the strong $q$ dependence of generalized Hurst
exponents and $\tau(q)$\cite{bun02}. The $q$- dependence of the
classical multifractal scaling exponent $\tau(q)$ has different
behaviors for $q<0$ and $q>0$. For positive and negative values of
$q$, the slopes of $\tau(q)$ are $0.45\pm0.02$ and $1.17\pm0.02$,
respectively. According to the value of Hurst exponent,
$H=0.52\pm0.01$, we find that this series has approximately random
behavior. This result is equal to value of Hurst exponent in small
scale of MF-DFA1 of noise with sinusoidal trend. The values of
derived quantities from MF-DFA1 method, are given in Table
\ref{Tab2}. Using the SWV we also analysis the truncated series, Our
result shows that the value of Hurst exponent is $H=0.50\pm0.01$,
which is in agreement with the previous result.

Usually, in the MF-DFA method, deviation from a straight line in the
log-log plot of Eq.~(\ref{Hq}) occurs for small scales $s$. This
deviation limits the capability of DFA to determine the correct
correlation behavior for very short scales and in the regime of
small $s$. The modified MF-DFA is defined as follows \cite{physa}:

\begin{eqnarray} F^{\rm mod}_q(s) &=& \frac{F_{q}(s)}{K_{q}(s)},\nonumber\\
& =& F_q(s) {\langle [F_q^{\rm shuf}(s')]^2 \rangle^{1/2} \, s^{1/2}
\over \langle [F_q^{\rm shuf}(s)]^2 \rangle^{1/2} \,
s'^{1/2} } \quad {\rm (for} \, s' \gg 1),\nonumber\\
\label{fmod}\end{eqnarray}
 where $\langle [F_q^{\rm shuf}(s)]^2
\rangle^{1/2}$ denotes the usual MF-DFA fluctuation function
[defined in Eq.~(\ref{fdef})] averaged over several configurations
of shuffled data taken from the original time series, and $s'
\approx N/40$. The values of the Hurst exponent obtained by modified
MF-DFA1 methods for duct water of Daugava river series is
$0.54\pm0.02$. The relative deviation of the Hurst exponent which is
obtained by modified MF-DFA1 in comparison to MF-DFA1 for original
data is approximately $3.84\%$. Now the value of Hurst exponent
ensure us that the runoff fluctuations are long-range correlation,
so by ignoring the seasonal trend, these processes have almost
memory. This means that the status of runoff water statistically has
long memory.

\section{Comparison of the multifractality for original,
shuffled and surrogate series}

As discussed in the section III the remanning data set after the
elimination of sinusoidal trend has the multifractal nature. In this
section we are interested in to determine the source of
multifractality.
 In general, two different
types of multifractality in time series can be distinguished: (i)
Multifractality due to a fatness of probability density function
(PDF) of the time series. In this case the multifractality cannot be
removed by shuffling the series. (ii) Multifractality due to
different correlations in  small and large scale fluctuations. In
this case the data may have a PDF with finite moments, e.~g. a
Gaussian distribution. Thus the corresponding shuffled time series
will exhibit mono-fractal scaling, since all long-range correlations
are destroyed by the shuffling procedure. If both kinds of
multifractality are present, the shuffled series will show weaker
multifractality than the original series. The easiest way to clarify
the type of multifractality, is by analyzing the corresponding
shuffled and surrogate time series. The shuffling of time series
destroys the long range correlation, Therefore if the
multifractality only belongs to the long range correlation, we
should find $h_{\rm shuf}(q) = 0.5$. The multifractality nature due
to the fatness of the PDF signals is not affected by the shuffling
procedure. On the other hand, to determine the multifractality due
to the  broadness of PDF, the phase of discrete fourier transform
(DFT) coefficients of the duct water of Daugava river time series
are replaced with a set of pseudo independent distributed uniform
$(-\pi,\pi)$ quantities in the surrogate method. The correlations in
the surrogate series do not change, but the probability function
changes to the Gaussian distribution. If multifractality in the time
series is due to a broad PDF, $h(q)$ obtained by the surrogate
method  will be independent of $q$. If both kinds of multifractality
are present in the duct water of Daugava river time series, the
shuffled and surrogate series will show weaker multifractality than
the original one.

\begin{figure}[t]
\epsfxsize=8truecm\epsfbox{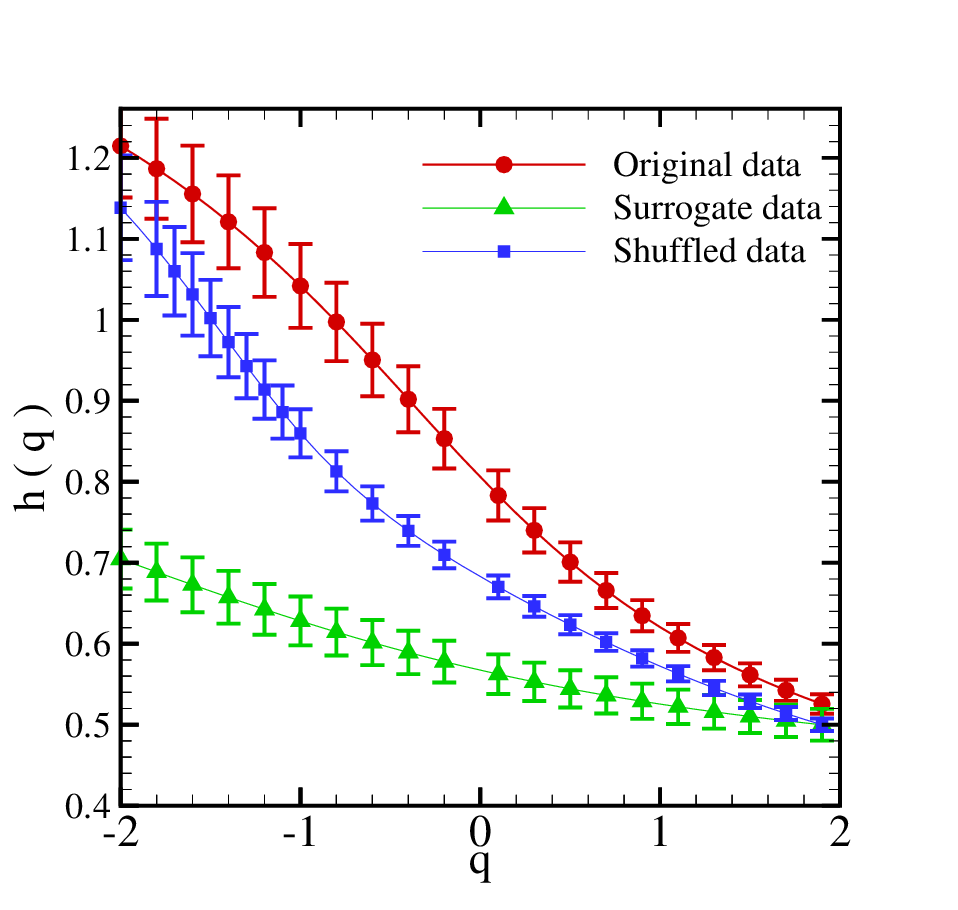} \narrowtext
\caption{Generalized Hurst exponent, $h(q)$ as a function of $q$ for
 original, surrogate and shuffled data.}
\label{fig6}
 \end{figure}

\begin{figure}[t]
\epsfxsize=8truecm\epsfbox{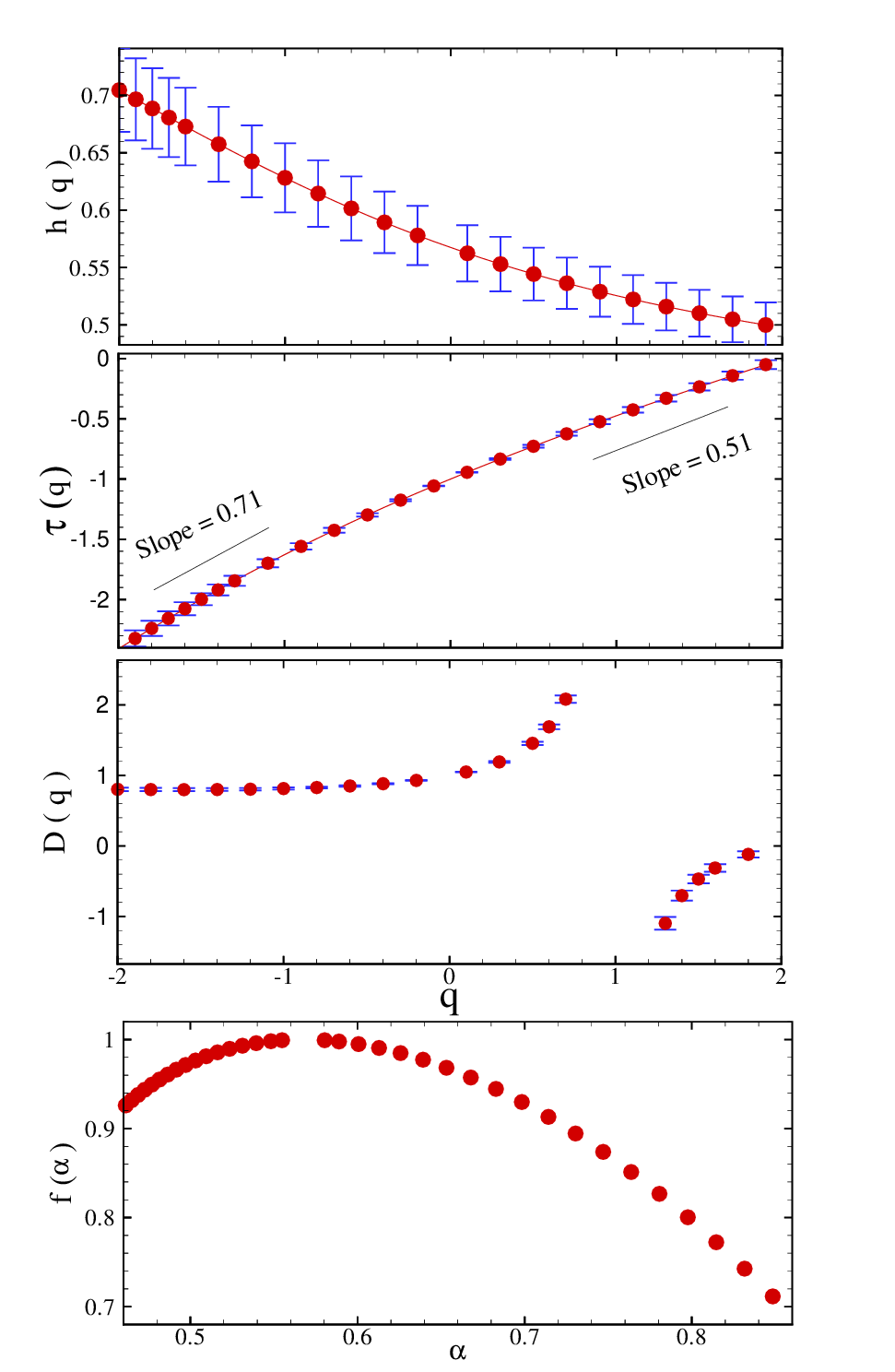} \narrowtext \caption{The $q$
dependence of the generalized Hurst exponent $h(q)$, the
corresponding $\tau(q)$, generalized multifractal dimension $D(q)$
and singularity spectrum $f(\alpha)$  are shown in the upper to
lower panel respectively for surrogate duct water of Daugava river
series without sinusoidal trend.} \label{fig7}
 \end{figure}

\begin{figure}[t]
\epsfxsize=8truecm\epsfbox{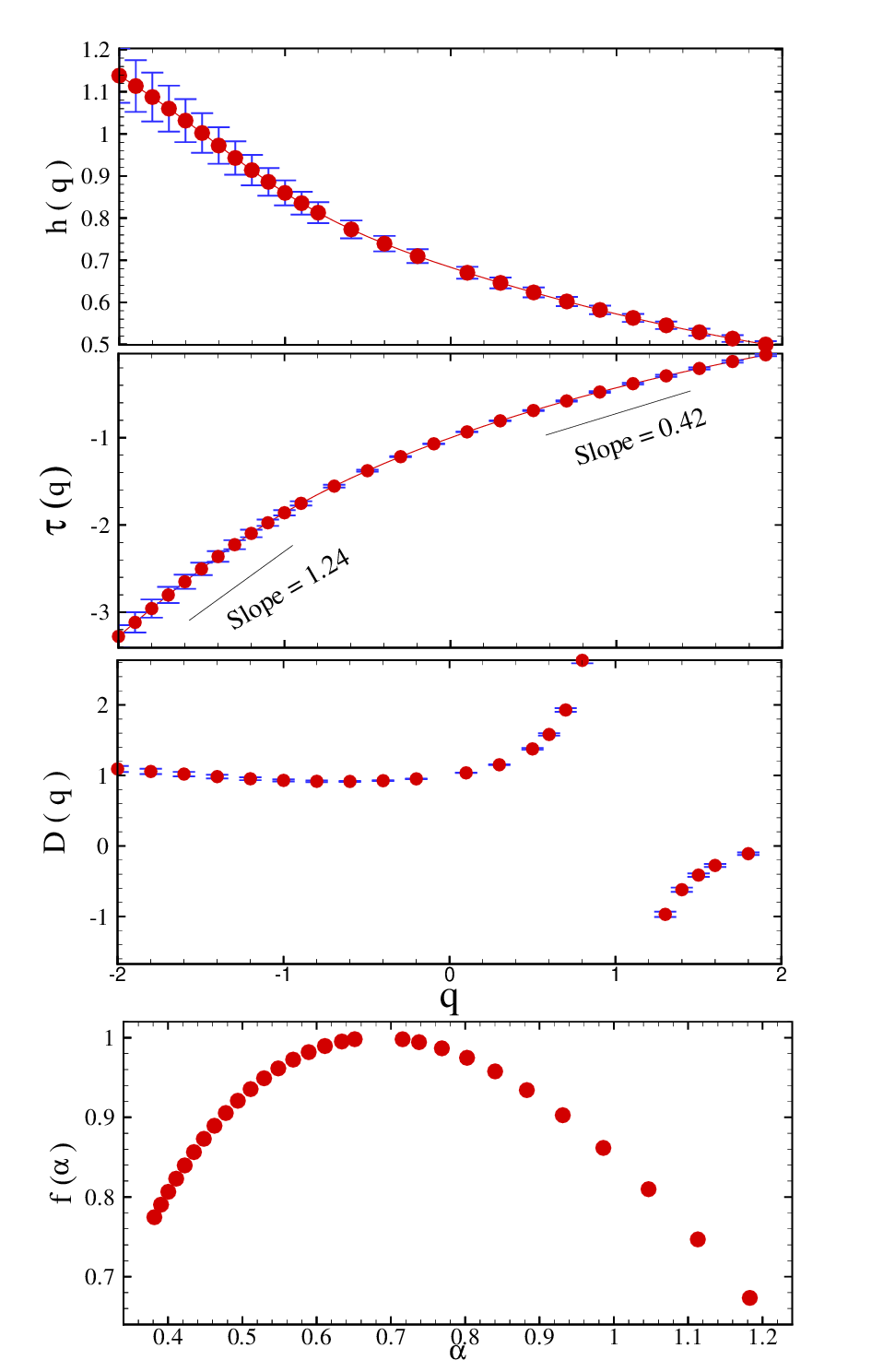} \narrowtext \caption{The $q$
dependence of the generalized Hurst exponent $h(q)$, the
corresponding $\tau(q)$, generalized multifractal dimension $D(q)$
and singularity spectrum $f(\alpha)$  are shown in the upper to
lower panel respectively for shuffled duct water of Daugava river
series without sinusoidal trend.} \label{fig8}
 \end{figure}

To check the nature of multifractality,  we compare the fluctuation
function $F_q(s)$, for the original series ( after cancelation of
sinusoidal trend) with the result of the corresponding shuffled,
$F_q^{\rm shuf}(s)$ and surrogate series $F_q^{\rm sur}(s)$.
Differences between these two fluctuation functions with the
original one, directly indicate the presence of long range
correlations or broadness of probability density function in the
original series. These differences can be observed in a plot of the
ratio $F_q(s) / F_q^{\rm shuf}(s)$ and $F_q(s) / F_q^{\rm sur}(s)$
versus $s$ \cite{bun02}. Since the anomalous scaling due to a broad
probability density affects $F_q(s)$ and $F_q^{\rm shuf}(s)$ in the
same way, only multifractality due to correlations will be observed
in $F_q(s) / F_q^{\rm shuf}(s)$. The scaling behavior of these
ratios are
\begin{equation} F_q(s) / F_q^{\rm shuf}(s) \sim s^{h(q)-h_{\rm
shuf}(q)} = s^{h_{\rm cor}(q)}, \label{HqCor} \end{equation}
\begin{equation} F_q(s) / F_q^{\rm sur}(s) \sim s^{h(q)-h_{\rm
sur}(q)} = s^{h_{\rm PDF}(q)}. \label{Hqpdf} \end{equation} If only
fatness of the PDF is responsible for the multifractality, one
should find $h(q)=h_{\rm shuf}(q)$ and $h_{\rm cor}(q)=0$. On the
other hand, deviations from $h_{\rm cor}(q) =0$ indicates the
presence of correlations, and $q$ dependence of $h_{\rm cor}(q)$
indicates that multifractality is due to the long rage correlation.
If only correlation multifractality is present, one finds $h_{\rm
shuf}(q)=0.5$. If both distribution and correlation multifractality
are present, both, $h_{\rm shuf}(q)$ and $h_{\rm sur}(q)$ will
depend on $q$. The $q$ dependence of the exponent $h(q)$ for
original, surrogate
 and shuffled time series are shown in
 Figure \ref{fig6}. The MF-DFA1 results of the surrogate and shuffled signal are shown in
 Figures
\ref{fig7} and \ref{fig8}, respectively. The $q$ dependence of
$h_{\rm cor}$ and
 $h_{\rm PDF}$ shows that the multifractality nature of the duct water of
Daugava river time series is due
 to both broadness of the  PDF and long range correlation. The absolute value of $h_{\rm cor}(q)$ is
 greater than $h_{\rm PDF}(q)$, so the multifractality due to the fatness
 is weaker than the multifractality due to the correlation.
 The deviation of $h_{\rm sur}(q)$ and $h_{\rm shuf}(q)$
 from $h(q)$ can be determined by using $\chi^2$ test as follows:
\begin{equation}
 \chi^2_{\diamond}=\sum_{i=1}^{N}
\frac{[h(q_i)-h_{\diamond}(q_i)]^2}{\sigma(q_i)^2+\sigma_{\diamond}(q_i)^2},
 \label{khi} \end{equation}
the symbol $"\diamond"$ can be replaced by $"\rm sur"$ and $"\rm
shuf"$, to determine the confidence level of $h_{\rm sur}$ and
$h_{\rm shuf}$ to generalized Hurst exponents of original series,
respectively. The value of reduced chi-square
$\chi^2_{\nu\diamond}=\frac{\chi^2_{\diamond}}{\cal{N}}$ ($\cal{N}$
is the number of degree of freedom) for shuffled and surrogate time
series are $6.98$, $29.97$, respectively. On the other hand the
width of singularity spectrum, $f(\alpha)$, i.e. $\Delta
\alpha=\alpha(q_{{\rm min}})-\alpha(q_{{\rm max}})$ for original,
surrogate and shuffled time series are approximately, $1.086$,
$0.390$ and $1.210$ respectively. These values also show that the
multifractality due to the broadness of the probability density
function is dominant\cite{paw05}.

The values of the generalized Hurst exponent $h(q=2.0)$,
multifractal scaling $\tau(q=2)$ and other related scaling exponents
are indicated in Table \ref{Tab2} for the original, shuffled of
 duct water of Daugava river series obtained with MF-DFA1 method are reported in
Table \ref{Tab2}. The values of the Hurst exponent obtained by
MF-DFA1 and modified MF-DFA1 methods for original, surrogate and
shuffled duct water of Daugava river series are given in Table
\ref{Tab3}.

\section{Conclusion}
The MF-DFA method allows us to determine the multifractal
characterization of the nonstationary and stationary time series. We
apply the recent method to investigate the existence of crossover on
the result of MF-DFA of river flow fluctuations. The concept of
MF-DFA of runoff water of rivers can be used to gain deeper insight
in to the processes occurring in climate and hydrological systems.
We have shown that the MF-DFA1 result of the monthly river flows
(e.g. duct water of Daugava river series) has two crossover time
scale $(s_{\times})$. Our results show that there exists a tendency
between some runoff river and sun activity. Indeed due to the
presence of  both seasonal trend and the effect of sun's period on
some river flows, we see at least two crossover in the DFA result of
rivers, nevertheless, it is not expected that the number of sunspots
has superior effect on runoff water fluctuations. So this effect has
different intensity on the rivers.
These crossover time scale are due to
the sinusoidal trend. To minimizing the effect of this trend, we
have applied F-DFA on the river flow time series. Applying the
MF-DFA1 method on truncated data, demonstrated that the monthly duct
water of Daugava river series is a stationary time series with
approximately random behavior. For other rivers we found long-range
correlation in their statistical behaviors. The $q$ dependence of
$h(q)$ and $\tau(q)$, indicated that the monthly duct water of
Daugava river series has multifractal behavior.  By comparing the
generalized Hurst exponent of the original time series with the
shuffled and surrogate one's, we have found that multifractality due
to the broadness of the probability density function has more
contribution than the correlation. The value of the Hurst exponent
shows that the flow of water without seasonal trend is the same as
random process.

{\bf Acknowledgements} We would like to thank  Mr. Rahimi Tabar and
Mr. Jafari for reading the manuscript and useful comments. This
paper is dedicated to Mrs. Somayeh Abdollahi and my son S. Danial
Movahed.
\begin{table}
\begin{center}
\caption{\label{Tab2}The values of the Hurst $(H)$, power spectrum
scaling $(\beta)$, auto-correlation scaling $(\gamma)$ exponents and
multifractal scaling for original, surrogate and shuffled of monthly
duct water of Daugava river series obtained by MF-DFA1.}
\begin{tabular}{|c|c|c|c|c|}
    Data & $H$ & $\beta$&$\gamma$&$\tau$\\ \hline
   Original & $0.52\pm 0.02$ &$0.04\pm0.02$ &$0.96\pm0.02$&$0.04\pm0.02$     \\\hline
   Surrogate &$0.51\pm0.02$ &$0.02\pm0.02$&$0.98\pm0.02$& $0.02\pm0.02$    \\\hline
  Shuffled & $0.50\pm0.01$ &$0.00$& $1.00\pm0.02$&$0.00$  \\ 
\end{tabular}
\end{center}
\end{table}
\begin{table}
\begin{center}
\caption{\label{Tab3}The value of the Hurst exponent using MF-DFA1
and modified MF-DFA1 for the original, shuffled and surrogate of
monthly duct water of Daugava river series.}
\begin{tabular}{|c|c|c|c|}
     Method & Original data & Surrogate &Shuffled \\ \hline
    MF-DFA1& $0.52\pm 0.01$ &$0.51\pm0.01$  &$0.50\pm0.01$     \\\hline
    Modified &$0.54\pm0.02$ &$0.46\pm0.02$ &$0.51\pm0.02$    \\
 \end{tabular}
\end{center}
\end{table}
\section{APPENDIX}
In this appendix we derive the relation between the exponent $h(2)$
(DFA1 exponent) and Hurst exponent of a fGn signal in one dimension.
We show that for such stationary signal the average sample variance
(Eq. \ref{fdef}) for $q=2$, is proportional to $s^{h(q)}$, where
$h(q=2)=H$. It is shown that the averaged sample variance $F^2(s)$
behaves as:
\begin{eqnarray}
F^2(s)&\equiv&  \frac{1}{N_{s}}\sum_{\nu=1}^{N_{s}} \left
[F^2(s;\nu) \right],\nonumber\\
&=&\left\langle \left [F^2(s;\nu)
\right]\right \rangle_{\nu},\nonumber\\
&\equiv& {\mathcal{C}_{H}}s^{2H},\label{ap2311}
\end{eqnarray}
where $F^2(s;\nu)$ is defined as:
\begin{eqnarray}
F^2(s;\nu)= \frac{1}{s}\sum_{i=1}^{s}\left [Y_{\nu}(i)-y_{\nu}(i)
\right]^2,\label{ap211}
\end{eqnarray}
and ${\mathcal{C}_{H}}$ is a function of Hurst exponent $H$. To
prove the statement we note that the data $x(k)$ is a fractional
Gaussian noise (fGn), the partial sums $Y(i)$ (Eq. \ref{profile})
will be a fBm signal:
\begin{eqnarray}
Y(i)=\sum_{k=1}^{i}x(k)-\langle x\rangle.
\end{eqnarray}

 In the DFA1, the fitting function
will have the expression $(y_{\nu}(i)=a_{\nu}+b_{\nu}i)$. The slope
$b_{\nu}$ and intercept $a_{\nu}$ of a least-squares line $Y(i)$ for
every windows $(\nu)$ are given by:
\begin{eqnarray}
b_{\nu}&=&\frac{\sum_{i=1}^{s}Y(i)i-\frac{1}{s}\sum_{i=1}^{s}Y(i)\sum_{i=1}^{s}i
}{\sum_{i=1}^{s}i^2-\frac{1}{s}\left[\sum_{i=1}^{s}i\right]^2},\nonumber\\
&&\simeq \frac{\sum_{i=1}^{s}Y(i)i-\frac{s}{2}\sum_{i=1}^{s}Y(i,j)
}{s^3/12},\nonumber\\
a_{\nu}&=&\frac{1}{s}\sum_{i=1}^{s}Y(i)-\frac{b_{\nu}}{s}\sum_{i=1}^{s}i,\nonumber\\&&\simeq\frac{1}{s}\sum_{i=1}^{s}Y(i)-\frac{b_{\nu}s}{2},
\label{ap12} \end{eqnarray} respectively.

 Using the Eqs. \ref{fdef} and \ref{ap12}, the Eq. \ref{ap211} can be written as
 follows:
\begin{eqnarray}
&&\left \langle\left [F^2(s;\nu) \right]\right\rangle =\left\langle
\frac{1}{s}\sum_{i=1}^{s}[Y(i)-a-bi]^2\right \rangle,\nonumber\\
&\simeq&\left\langle
\frac{1}{s}\sum_{i=1}^{s}Y(i)^2\right\rangle+\left\langle
a^2\right\rangle+\frac{s^2}{3}\left\langle
b^2\right\rangle+s\left\langle ab
\right\rangle\nonumber\\
&&-2\left\langle
\frac{a}{s}\sum_{i=1}^{s}Y(i)\right\rangle-2\left\langle
\frac{b}{s}\sum_{i=1}^{s}iY(i)\right\rangle,
 \label{a33}
\end{eqnarray}
where we have discard the subscript $\nu$ for simplicity.
 The fBm signals is produced by using the fGn noise as follows:
\begin{eqnarray}
Y(i)=i^Hx, \label{a4}
\end{eqnarray}
 and
\begin{eqnarray}
Y(i)-Y(k)&=&|i-k|^Hx,
 \end{eqnarray}
 so,
\begin{eqnarray}
\langle[Y(i)-Y(k)]^2\rangle=\sigma^2|i-k|^{2H},\label{23}
\end{eqnarray}
where $\sigma^2=\left \langle x(i)^2\right\rangle$. The variance of
fBm signal is: $\left \langle
Y(i)^2\right\rangle=\sigma^2i^{2H}$\cite{Peng94}. Expanding the left
hand side of Eq. \ref{23}, it can be easily shown that the
correlation function of $Y(i)$, has the following form:
\begin{eqnarray}
\left\langle
Y(i)Y(k)\right\rangle&=&\frac{\sigma^2}{2}[i^{2H}+k^{2H}-|i-k|^{2H}].
\label{a55}
\end{eqnarray}

 Finally using the Eqs. \ref{ap12} and  \ref{a55}, it can be easily
shown that the Eq. \ref{a33} can be written as follows:
\begin{eqnarray}
\left \langle\left [F^2(s;\nu)
\right]\right\rangle_{\nu}={\mathcal{C}_{H}}(s)^{2H}. \label{a66}
\end{eqnarray}

To determine the ${\mathcal{C}_{H}}$, we have to calculate some
terms such as:
\begin{eqnarray}
&&\sum_{i,j=1}^{s}\langle
iY(i)Y(j)\rangle=\frac{\sigma^2}{2}\sum_{i,j=1}^{s}\left(i^{2H+1}+ij^{2H}-i|i-j|^{2H}\right),\nonumber\\
&&=\frac{\sigma^2}{2}\sum_{i,j=1}^{s}\left(i^{2H+1}+ij^{2H}\right)
-\frac{\sigma^2}{2}\sum_{i=1}^{s}\sum_{j=1}^{i}i(i-j)^{2H}\nonumber\\&&
\qquad-\frac{\sigma^2}{2}\sum_{i=1}^{s}\sum_{j=i}^{s}i(j-i)^{2H},\nonumber\\
&&\sim\frac{\sigma^2}{2}\left(\frac{s^{2H+3}}{2H+2}+\frac{s^{2H+3}}{2(2H+1)}\right)\nonumber\\
&&-\frac{\sigma^2}{2}\sum_{i=1}^{s}i^{2H+2}\left(\int_{0}^{1}(1-x)^{2H}dx-\int_{0}^{1}x(1-x)^{2H}dx\right),\nonumber\\
&&=\frac{\sigma^2s^{2H+3}}{4}\left(\frac{2}{H+1}-\frac{1}{2H+1}\right),
\end{eqnarray}
and
\begin{eqnarray}
&&\sum_{i,j=1}^{s}\langle Y(i)Y(j)\rangle=\frac{\sigma^2}{2}\sum_{i,j=1}^{s}\left(i^{2H}+j^{2H}-|i-j|^{2H}\right),\nonumber\\
&&=\frac{\sigma^2}{2}\sum_{i,j=1}^{s}\left(i^{2H}+j^{2H}\right)-\sigma^2\sum_{i=1}^{s}\sum_{j=1}^{i}(i-j)^{2H},\nonumber\\
&&\sim\sigma^2\left(\frac{s^{2H+2}}{2H+1}-\sum_{i=1}^{s}i^{2H+1}\int_0^1(1-x)^{2H}\right),\nonumber\\
&&\sim\sigma^2s^{2H+2}\left(\frac{1}{2H+1}-\frac{1}{(2H+2)(2H+1)}\right).
\end{eqnarray}

Finally ${\mathcal{C}_{H}}$ has the following form:
\begin{eqnarray}
{\mathcal{C}_{H}}&=&\frac{\sigma^2}{(2H+1)}-\frac{4\sigma^2}{2H+2}\nonumber\\
&&+3\sigma^2\left(\frac{2}{H+1}-\frac{1}{2H+1}\right)\nonumber\\
&&\qquad-\frac{3\sigma^2}{(H+1)}\left(1-\frac{1}{(H+1)(2H+1)}\right).
\label{a77}
\end{eqnarray}

Therefore the standard DFA$1$ exponent for a stationary signal is
related to its Hurst exponent as $h(q=2)=H$.


\begin{thebibliography} {50}
\bibitem{note3}  For the maximum scale $s=N$ the fluctuation function
$F_q(s)$ is independent of $q$, since the sum in Eq.~(\ref{fdef})
runs over only two identical segments ($N_s \equiv [N/s] = 1$). For
smaller scales $s \ll N$ the averaging procedure runs over several
segments, and the average value $F_q(s)$ will be dominated by the
$F^2(s,\nu)$ from the segments with small (large) fluctuations if $q
< 0$ ($q > 0$).  Thus, for $s \ll N$, $F_q(s)$ with $q < 0$ will be
smaller than $F_q(s)$ with $q > 0$, while both become equal for
$s=N$.  Hence, if we assume an homogeneous scaling behavior of
$F_q(s)$ following Eq.~(\ref{Hq}), the slope $h(q)$ in a log-log
plot of $F_q(s)$ with $q < 0$ versus $s$ must be larger than the
corresponding slope for $F_q(s)$ with $q > 0$. Thus, $h(q)$ for $q <
0$ will usually be larger than $h(q)$ for $q
> 0$.
\end{thebibliography}
\end{document}